\documentclass[a4paper,11pt]{article}
\pdfoutput=1 
\usepackage{jcappub, bm, color} 
\usepackage{amssymb,amsfonts,slashed,amsthm,amsmath,graphicx, soul, empheq}
\usepackage[caption=false]{subfig}
\begin{document}

\renewcommand{\figurename}{Fig.}
\renewcommand{\tablename}{Table.}
\newcommand{\Slash}[1]{{\ooalign{\hfil#1\hfil\crcr\raise.167ex\hbox{/}}}}
\newcommand{\bra}[1]{ \langle {#1} | }
\newcommand{\ket}[1]{ | {#1} \rangle }
\newcommand{\beq}{\begin{equation}}  \newcommand{\eeq}{\end{equation}}
\newcommand{\bef}{\begin{figure}}  \newcommand{\eef}{\end{figure}}
\newcommand{\bec}{\begin{center}}  \newcommand{\eec}{\end{center}}
\newcommand{\non}{\nonumber}  \newcommand{\eqn}[1]{\begin{equation} {#1}\end{equation}}
\newcommand{\laq}[1]{\label{eq:#1}}  
\newcommand{\dd}[1]{{d \o d{#1}}}
\newcommand{\Eq}[1]{Eq.~(\ref{eq:#1})}
\newcommand{\Eqs}[1]{Eqs.~(\ref{eq:#1})}
\newcommand{\eq}[1]{(\ref{eq:#1})}
\newcommand{\Sec}[1]{Sec.\ref{chap:#1}}
\newcommand{\ab}[1]{\left|{#1}\right|}
\newcommand{\vev}[1]{ \left\langle {#1} \right\rangle }
\newcommand{\bs}[1]{ {\boldsymbol {#1}} }
\newcommand{\lac}[1]{\label{chap:#1}}
\newcommand{\SU}[1]{{\rm SU{#1} } }
\newcommand{\SO}[1]{{\rm SO{#1}} }
\def\({\left(}
\def\){\right)}
\def\dt{{d \o dt}}
\def\diag{\mathop{\rm diag}\nolimits}
\def\Spin{\mathop{\rm Spin}}
\def\O{\mathcal{O}}
\def\U{\mathop{\rm U}}
\def\Sp{\mathop{\rm Sp}}
\def\SL{\mathop{\rm SL}}
\def\tr{\mathop{\rm tr}}
\def\ebq{\end{equation} \begin{equation}}
\newcommand{\OR}{~{\rm or}~}
\newcommand{\AND}{~{\rm and}~}
\newcommand{\EV}{ {\rm \, eV} }
\newcommand{\KEV}{ {\rm \, keV} }
\newcommand{\MEV}{ {\rm \, MeV} }
\newcommand{\GEV}{ {\rm \, GeV} }
\newcommand{\TEV}{ {\rm \, TeV} }
\def\o{\over}
\def\a{\alpha}
\def\b{\beta}
\def\c{\varepsilon}
\def\d{\delta}
\def\e{\epsilon}
\def\f{\phi}
\def\g{\gamma}
\def\h{\theta}
\def\k{\kappa}
\def\l{\lambda}
\def\m{\mu}
\def\n{\nu}
\def\p{\psi}
\def\q{\partial}
\def\r{\rho}
\def\s{\sigma}
\def\t{\tau}
\def\u{\upsilon}
\def\v{\varphi}
\def\w{\omega}
\def\x{\xi}
\def\y{\eta}
\def\z{\zeta}
\def\D{\Delta}
\def\G{\Gamma}
\def\H{\Theta}
\def\L{\Lambda}
\def\F{\Phi}
\def\P{\Psi}
\def\S{\Sigma}
\def\me{\mathrm e}
\def\ol{\overline}
\def\tl{\tilde}
\def\*{\dagger}



\begin{flushright}
TU-1124
\end{flushright}

\title{
Challenges for heavy QCD axion inflation
}

\author{
Fuminobu Takahashi,
Wen Yin
}
\affiliation{
Department of Physics, Tohoku University, 
Sendai, Miyagi 980-8578, Japan
}

\abstract{
We examine the theoretical possibility for the heavy QCD axion to induce slow-roll inflation while solving the strong CP problem through the Peccei-Quinn mechanism. If the cancellation between contributions from a small-size instanton and another Peccei-Quinn symmetry breaking  occurs with high accuracy, the potential can be flattened enough near its maximum to achieve hilltop inflation. This comes at the cost of severe fine-tuning of their relative size and phase, but it also allows us to relate the high quality problem of the Peccei-Quinn symmetry to the fine-tuning problem of the hilltop inflation.
There are two classes of such axion hilltop inflation, each giving a different relation between the axion mass at the minimum and the decay constant. 
 The first class predicts the relation $m_\phi \sim 10^{-6}f_\phi$, and the axion can decay via the gluon coupling 
and reheat the universe. 
Most of the predicted parameter region will be covered by various experiments such as  CODEX, DUNE, FASER, LHC, MATHUSLA, and NA62 where
the production and decay proceed through the same coupling that induced reheating. The second class predicts the relation 
$m_\phi \sim 10^{-6} f^2_\phi/M_{\rm pl}$. In this case, the axion mass is much lighter than in the previous case, and one needs another
mechanism for successful reheating. The viable decay constant is restricted to be $10^8\GEV \lesssim f_\phi \lesssim 10^{10}\GEV$, which will be
probed by future experiments on the electric dipole moment of nucleons. 
In both cases, requiring the axion hilltop inflation results in the strong CP phase that is close to zero.
}

\emailAdd{fumi@tohoku.ac.jp, yin.wen.b3@tohoku.ac.jp}

\maketitle
\flushbottom

\section{Introduction}
The strong CP problem is one of the unsolved problems in the standard model (SM), and it strongly suggests the existence of new physics beyond the SM.
One of the solutions is the Peccei-Quinn (PQ) mechanism~\cite{Peccei:1977hh,Peccei:1977ur}.
This mechanism predicts a light pseudo-Nambu-Goldstone boson, called the QCD axion, whose dynamics adjusts the strong CP phase $\bar{\theta}_{\rm QCD}$ to zero~\cite{Weinberg:1977ma,Wilczek:1977pj}. 
In the standard QCD axion scenario, its mass is mainly generated by the non-perturbative effects of QCD, and it must be lighter than ${\cal O}(10)$ meV to satisfy astrophysical 
bounds~\cite{Hamaguchi:2018oqw,Beznogov:2018fda,Leinson:2019cqv}. See also Refs.\,\cite{Jaeckel:2010ni,Ringwald:2012hr,Arias:2012az,Graham:2015ouw,Marsh:2015xka,Irastorza:2018dyq, DiLuzio:2020wdo} for reviews.

While the QCD axion is a good candidate for dark matter, it is not considered to have driven inflation.
The only free parameter of the QCD axion potential is the decay constant. Since the height of the potential is about the QCD scale, no matter how we change the decay constant, we cannot satisfy both the slow-roll conditions and the CMB normalization. Therefore,  to identify the QCD axion with the inflaton, we need to extend the axion sector in some way, for example, by introducing extra potential terms. However, such extra potential terms required to realize inflation largely break the PQ symmetry and they may spoil the PQ mechanism. In this Letter we investigate whether this is indeed the case, and clarify the theoretical challenges involved.

In recent years, there has been a lot of interest in scenarios where the QCD axion is much heavier than what is considered in the standard scenario, often referred to as the ``heavy QCD axion." There are various ways to increase the mass of the QCD axion without spoiling the PQ mechanism.  To do so, the extra potential must have a minimum  at or very near  $\bar{\theta}_{\rm QCD} = 0$. In other words, the extra potential must be aligned to that from the non-perturbative QCD effects.
As such, small-size instantons have been studied in the context of the PQ quality problem and increasing the QCD axion mass~\cite{Holdom:1982ex, Dine:1986bg, Flynn:1987rs, Agrawal:2017ksf, Csaki:2019vte,Gherghetta:2020ofz,Gupta:2020vxb,Poppitz:2002ac,Gherghetta:2020keg}. Interestingly, it was recently shown that the potential from the small-size instantons can be aligned to that from the IR QCD instanton in several setups~\cite{Agrawal:2017ksf, Csaki:2019vte,Gherghetta:2020ofz,Gupta:2020vxb,Poppitz:2002ac,Gherghetta:2020keg}, as long as 
the higher dimensional terms of the SM fermions are small enough at the instanton scale \cite{Kitano:2021fdl}. 
Alternatively, it was discussed in Refs.\,\cite{Rubakov:1997vp,Berezhiani:2000gh, Gianfagna:2004je,Hsu:2004mf,Hook:2014cda,Fukuda:2015ana,Gherghetta:2016fhp} that another strong dynamics related to the QCD may provide a 
heavy mass for the QCD axion.

In this Letter, 
 we consider the possibility of achieving slow-roll inflation in a heavy QCD axion scenario, and identify what theoretical difficulties it entails. 
In general, an axion is known to be a good candidate for the inflaton, since its potential is protected from large radiative corrections by the discrete shift symmetry,
\begin{eqnarray}
\laq{shift}
\phi  \rightarrow \phi + 2  \pi f_\phi,
\end{eqnarray}
where $\phi$ is the axion, and $f_\phi$ is the decay constant.
The axion potential can be generated via non-perturbative effects originating from UV or IR dynamics. 
If the potential is dominated by a single instanton effect, then so-called natural inflation can be realized for a super-Planckian decay constant~\cite{Freese:1990rb,Adams:1992bn}.
Such a large decay constant could be in conflict with quantum gravity.
Moreover, the latest measurements of the anisotropy of the cosmic microwave background (CMB) strongly disfavor this model~\cite{Akrami:2018odb}. On the other hand, if the potential is dominated by two or more instanton contributions, successful inflation is possible with a sub-Planckian decay constant, and the inflationary scale can be much smaller. Such inflation is called multi-natural inflation~\cite{Czerny:2014wza, Czerny:2014xja,Czerny:2014qqa,Croon:2014dma,Higaki:2014sja}. We emphasize that this comes at the cost of fine tuning the parameters of various contributions, and the amount of fine-tuning is
of ${\cal O}(f_\phi^3/M_{\rm pl}^3)$, which becomes worse for smaller decay constant.\footnote{
The fine-tuning of the parameters required for successful slow-roll inflation may be
explained by anthropic argument.
}

Recently, Daido and the present authors studied 
cosmological and phenomenological aspects of the multi-natural inflation caused by an axion-like particle (ALP)~\cite{Daido:2017wwb, Daido:2017tbr,Takahashi:2019qmh}. The ALP is usually referred to as an axion that has a coupling with photons.
The most minimal multi-natural inflation is based on a potential with two instanton contributions. Then, inflation occurs around the top of the potential or at an inflection point, and the inflation can be
eternal only for the hilltop inflation~\cite{Takahashi:2019qmh}.
In the axion hilltop inflation, the models can be further divided into two classes. In the first class, at the potential minimum, the mass of the axion (inflaton) has the magnitude naturally expected from the height of the potential and the decay constant. Such axion can be searched for in accelerator experiments like the SHiP experiment.~\cite{Anelli:2015pba,Alekhin:2015byh,Dobrich:2015jyk, Dobrich:2019dxc}. 
The inflationary Hubble scale, $H_{\rm inf}$, within the SHiP reach is $H_{\rm inf} \lesssim 1\EV$.\footnote{
\label{ft1}
In the eternal inflation setup, a light scalar has a probability distribution peaked at the minimum due to the stochastic dynamics~\cite{Graham:2018jyp,Guth:2018hsa}. Then,
the cosmological moduli problem such as the overproduction of the light axions in the axiverse is absent if $H_{\rm inf}\lesssim 0.1\KEV$~\cite{Ho:2019ayl,Reig:2021ipa}.}
In the second class, the axion potential has an upside-down symmetry,
and the axion mass at the minimum is very light, and comparable to the Hubble parameter during inflation. In this case, the reheating is incomplete, and the remnant of the axion condensate becomes dark matter.
Thus,  it is possible to unify
the inflaton and dark matter in terms of a single ALP, the so-called ``the ALP miracle" scenario~\cite{Daido:2017wwb, Daido:2017tbr}. 
The viable parameter region can be fully tested in the next-generation axion helioscope, IAXO experiment~\cite{Irastorza:2011gs, Armengaud:2014gea, Armengaud:2019uso, Abeln:2020ywv} and future observations of 
CMB and baryonic acoustic oscillation~\cite{Kogut:2011xw, Abazajian:2016yjj, Baumann:2017lmt}.\footnote{Interestingly the ALP may explain the XENON1T excess \cite{Aprile:2020tmw, Takahashi:2020uio} if the PQ symmetry is anomaly free under the SM gauge groups \cite{Takahashi:2020bpq}.}  
Laser-based photon colliders~\cite{Hasebe:2015jxa,Fujii:2010is,Homma:2017cpa}, which has been already operated \cite{Homma:2021hnl}, can also be an alternative probe of this scenario.

In this Letter, we show that, in the heavy QCD axion scenario, the QCD axion can induce successful slow-roll inflation, at the cost of fine-tuning parameters of the potential terms. The inflaton dynamics itself is analogous to the hilltop inflation induced by an ALP, but it has two novel features. First, the extra potential required for having a sufficiently flat plateau around the top of the potential does not contribute to the strong CP phase. This is because the phase contribution is suppressed by the slow-roll condition for the inflaton. This consistency with the PQ solution is a unique feature of the axion hilltop inflation.
Second, the QCD axion has a gluonic coupling, and a large portion of the mass and coupling can be searched for in various experiments.
Our QCD axion inflation model predicts a certain relation between the mass and the decay constant as in the ALP inflation~\cite{Takahashi:2019qmh}, providing an interesting parameter region to be explored by those experiments. Also, if the QCD axion is relatively light, the contribution of the extra potential to the strong CP phase can be probed by the precise measurements of the nucleon electric dipole moment (EDM).

The rest of this Letter is organized as follows. We first review the axion hilltop inflation in Sec~\ref{sec:2}. The cosmological and experimental implications of the heavy QCD axion inflation is discussed in Sec.~\ref{sec:3}. The last section is devoted for discussion.

\section{Review of axion hilltop inflation}
\label{sec:2}
We focus on the minimal case of the multi-natural inflation, where the potential is dominated by the two cosine terms,
\begin{align}
\label{eq:DIV} 
V_{\rm inf}(\phi) = \Lambda^4\(\cos\(\frac{\phi}{f_\f} + \Theta \)- \frac{\kappa }{n^2}\cos\(n\frac{\f}{f_\f }\)\) + {\rm const.}
\end{align}
Here $n$ ($>1$) is an integer,\footnote{ \label{ft2}
One can consider a case in which the first cosine term in \Eq{DIV} contains
another positive integer $n' < n$. It is straightforward to extend our analysis to this case by redefining the decay constant. 
In particular, we have $|\f_{\rm max}/f_\phi -\f_{\rm min}/f_\f|\simeq \pi /n' \mod {2\pi}$ if $n/n'$ is an integer. In this case our conclusion still holds, if $c_g/n'$ is odd, where $c_g$ is defined in \eq{Lag}. If this is not the case, the strong CP phase would be generically of order unity.
} $\kappa$ and $\Theta$ parameterize the relative height and phase of the two terms, 
respectively. The constant term is to make the cosmological constant vanishingly small in the present vacuum. If the two terms conspire to make the top of the potential extremely flat, slow-roll inflation takes place around the hilltop. 
As we shall see shortly,
the parity of $n$  characterizes the shape of the potential and the relation between the ALP mass, $m_\f$, and decay constant, $f_\f$, at the vacuum.  See Fig.\,\ref{fig:pot}.

\begin{figure}[!t]
\begin{center}  
     \includegraphics
[width=145mm]{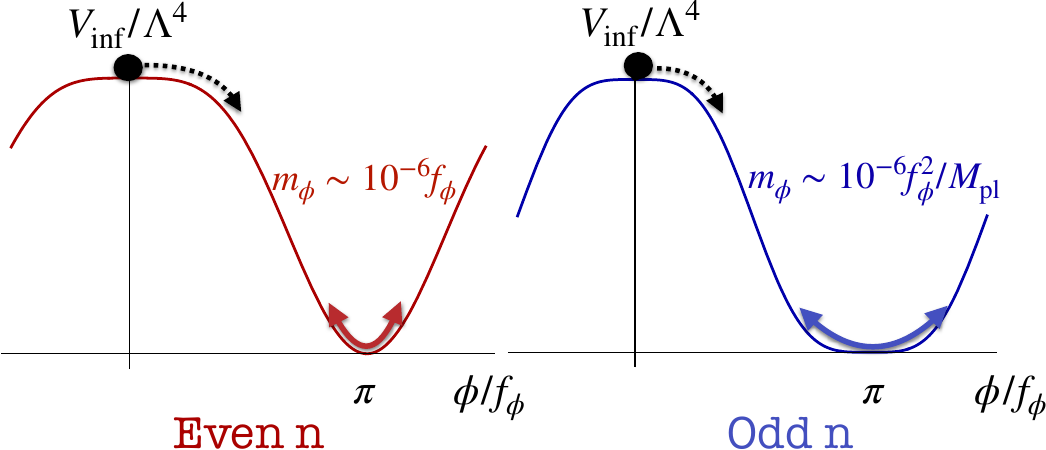}
      \end{center}
\caption{ The typical potential shape for even $n$ [left panel] and odd $n$ [right panel].  
The typical relations between mass, $m_\f,$ and decay constant, $f_\f,$  at around the vacuum are also shown. 
{We take $\Theta=0, \k=1$ for illustrative purpose.}
}
\label{fig:pot} 
\end{figure}

Let us briefly review the inflation dynamics with the potential \eq{DIV}, following Refs.~\cite{Daido:2017wwb, Daido:2017tbr, Takahashi:2019qmh}.
Let us expand the potential around $\f=0$, 
\begin{equation}
\label{eq:app}
V_{\rm inf}(\phi) \simeq V_0   - \Theta \frac{\L^4 }{ f_\f} {\f} +\frac{m^2 }{2}\f^2 - \lambda \f^4+\cdots,    
\end{equation}
where the dots include a cubic term and higher order ones that are irrelevant for the inflaton dynamics during inflation.
Here we have defined 
 \begin{align}
 \label{eq:V0}
 V_0 & \equiv \left(2 - \frac{2}{n^2} \sin^2{\frac{n \pi}{2}} \right) \Lambda^4,\\
 m^2 &\equiv {\(\kappa-1\)}\frac{\L^4 }{f_\f^2},\\
 \lambda &\equiv \frac{n^2-1 }{4!}\(\frac{\Lambda}{f_\f}\)^4.
 \label{eq:lambda}
\end{align}
The potential can be extremely flat around the hilltop  with certain $\k$ and $\Theta$. At the potential maximum, $\phi = \phi_{\rm max}$, $V_{\rm inf}'(\phi_{\rm max})$ vanishes, and if the curvature  satisfies $|V_{\rm inf}''(\phi_{\rm max})|\lesssim 
 V_{\rm inf}(\phi_{\rm max})/M_{\rm pl}^2 \simeq
 3 H_{\rm inf}^2$, the slow-roll inflation takes place.
 Here $M_{\rm pl} \simeq 2.4 \times 10^{18}\GEV$ is the reduced Planck mass.

One can see that the relative phase $\Theta$ leads to a  (tiny) linear term of $\phi$, i.e. the second term in \Eq{app}, which is necessary to explain the observed scalar spectral index, $n_s$. This is because the predicted value of $n_s$ in the quartic hilltop inflation with $\Theta = 0$ and $\kappa=1$ is known to be too small unless the field excursion of the inflaton becomes comparable to or larger than the Planck mass. The small linear term can effectively increase the predicted value of $n_s$~\cite{Takahashi:2013cxa}.\footnote{
The inflection point inflation cannot be eternal due to the linear term required to explain the observed spectral index~\cite{Takahashi:2019qmh}. The same argument was also used for the inflation model satisfying the trans-Planckian censorship conjecture~e.g. \cite{Kadota:2019dol}.}

We assume that $\kappa$ and $\Theta$ are in the following
range~\cite{Takahashi:2019qmh},
\beq
\label{cfei}
 |\k-1|\lesssim \(\frac{f_\f}{M_{\rm pl}}\)^2,~~|\Theta|\lesssim \(\frac{f_\f}{M_{\rm pl}}\)^3,
\eeq
for which the slow-roll conditions are satisfied at the potential maximum. 
If the prior probability distribution of $\kappa$ and $\Theta$
are flat in the relevant range, the smallness of the above parameters represents the amount of fine-tuning, which becomes more severe for smaller $f_\phi$.
Note that, if the axion hilltop inflation is eternal, it may be possible to partially compensate the tuning of the parameters.
The inflationary Hubble parameter is given by
\beq
H_{\rm inf} = \sqrt{\frac{V_0}{3M^2_{\rm pl}}}\sim \frac{\L^2}{M_{\rm pl}}.
\eeq
These conditions imply that the location of the potential maximum is determined by balancing the linear term with the quartic one, and it is at 
\beq
\laq{cmax}
\frac{\f_{\rm max}}{f_\f} \;\simeq\; -\(\frac{6\x}{n^2-1}\)^{1/3} \frac{f_\f}{M_{\rm pl}},
\eeq
where  $\xi$ is a positive parameter of ${\cal O}(0.01)$, which depends on 
 $\k\times  (f_\f/M_{\rm pl})^{-2}$ and $\Theta \times  (f_\f/M_{\rm pl})^{-3}$.
  Here and hereafter we assume that the slow-rolling inflaton moves toward the $\phi>0$ direction without loss of generality. 
Apart from the spectral index, this model is almost a quartic hilltop inflation, and the CMB normalization of primordial density perturbation fixes the quartic coupling, $\l$, or equivalently the ratio of $\Lambda$ and $f_\phi$ as~\cite{Akrami:2018odb,Takahashi:2019qmh}
\beq
\label{eq:ratio}
\frac{\Lambda}{f_\f}\simeq  1.2\times 10^{-3 }\(\frac{3}{n^2-1}\)^{\frac{1}{4}}\(\frac{30}{ N_*}\)^{\frac{3}{4}}.
\eeq
Here $N_*$ is the e-folding number at the horizon exit of the CMB scale.

The properties of the axion at the potential minimum such as the mass and decay constant have important implications for low-energy phenomena, and they are related to each other by the CMB observations. The axion hilltop inflation can be divided into two classes depending on the parity of $n$. 

When $n$ is even, the axion mass has the usual scaling $m_\phi \sim \Lambda^2/f_\phi$.  
By minimizing the potential we obtain 
\begin{align}
\laq{feven}
\vev{\f} \approx \pi f_\phi -\frac{\Theta}{2} f_\f 
,~~~~m_\phi \approx \sqrt{2}\frac{\L^2}{f_\phi}~~~ \text{[for even $n$]},
\end{align}
where we have used the \Eq{cmax}. Thus, we obtain the relation between $m_\phi$ and $f_\phi$ from the CMB normalization (\ref{eq:ratio}). See the left panel of Fig.~\ref{fig:pot}.

When $n$ is odd, on the other hand, the potential has an upside-down symmetry, 
\beq
V_{\rm inf}(\f)=-V_{\rm inf}(\f+\pi f_\f)+\text{const.}\eeq
The shape near the potential minimum is the same as the potential maximum turned upside down. See the right panel of Fig.~\ref{fig:pot}.
As a result, we obtain  
\begin{align}
\laq{fodd}
\vev{\f} = \pi f_\phi +{\f_{\rm max}},  
~~~~m_\phi = \sqrt{-V''_{\rm inf} (\f_{\rm max})}\sim H_{\rm inf}~~~ \text{[for odd $n$]},
\end{align}
In the last approximation we have used that $|V''_{\rm inf}(\phi_{\rm max})|\sim H_{\rm inf}^2$ for the (eternal) hilltop inflation.
Other than the approximation, the above equations hold exactly due to the upside-down symmetry. 
Here the nonzero mass reflects the fact that we have introduced a linear term to explain the spectral index.

\section{Heavy QCD axion inflation}
\label{sec:3}

\subsection{The strong CP problem solved 
by the axion inflation}
Here we first show that the axion hilltop inflation can be implemented in the heavy QCD axion scenario without spoiling the PQ mechanism, at the cost of fine tuning the parameters. We will come back to the issue of  fine-tuning later.

In the standard PQ mechanism, the potential of the QCD axion receives a contribution from the IR QCD effects so that the strong CP phase is dynamically set to zero at its potential minimum. In the heavy QCD axion scenario,  the axion has extra PQ breaking terms whose minima must be aligned with that of the potential from the IR QCD effects to solve the strong CP problem.

As a concrete example, let us consider a simple UV model based on  the extended gauge groups, $\SU(3)_1 \times \SU(3)_2$, studied in Ref.~\cite{Agrawal:2017ksf}.
If the axion is coupled to both $\SU(3)_1$ and  $\SU(3)_2$, its couplings in the UV are given by
\beq
{\cal L} \supset \frac{\a_1}{8\pi}\( c_1  \frac{\f}{f_\f}-\bar \theta_1\) G_1 \tl{G_1}+\frac{\a_2}{8\pi}\( c_2  \frac{\f}{f_\f}-\bar \theta_2\) G_2 \tl{G_2}
\eeq
where $G_{i}$, $\a_{i}$, $c_{i}$, and $\bar\theta_i$ are the field strength, coupling constant,  anomaly coefficient, and CP phase of the $\SU(3)_i\,(i=1,2)$, respectively. Note that this model does not solve the strong CP problem {by itself}, because unlike the original setup, here we consider only one axion coupled to both gauge groups.
If we introduce a bi-fundamental Higgs field which develops a nonzero expectation value,
the $\SU(3)_1 \times \SU(3)_2$ symmetry is spontaneously broken to its diagonal subgroup, which is identified with $\SU(3)_c$.
After integrating out the heavy degrees of freedom, 
we obtain the low-energy effective Lagrangian~\cite{Agrawal:2017ksf}
\beq
\laq{Lageff1}
{\cal L}_{\rm eff}=  \L_1^4  \cos\(c_1\frac{\f}{f_\f} -\bar \theta_1\)+\L_2^4 \cos\(c_2\frac{\f}{f_\f} -\bar \theta_2\)+\frac{\a_s}{8\pi}\(( c_1 +c_2 ) \frac{\f}{f_\f}-\bar \theta_1-\bar \theta_2\) G \tl{G}.
\eeq
Here the first two terms are from the small-size instantons,
and the last term represents the axion coupling to gluons. 
What is interesting is that once we require that the axion hilltop inflation should take place in this set-up, the relative height and phase of the first two terms must be aligned, and as a result, the CP symmetry is conserved at the potential minimum. 
In Appendix~\ref{chap:app1} we provide a possible UV completion with two axions. 

As in the above example, the relevant low-energy Lagrangian boils 
down to the following one,
\begin{align}
\laq{Lag}
{\cal L}_{\rm eff}= 
-\L^4  \(\cos\(\frac{\f}{f_\f} +\H\)-\frac{\k}{n^2} \cos\(n \frac{\f}{f_\f}  \)\)+\frac{\a_s}{8\pi}\( c_g \frac{\f}{f_\f}-\pi+\H\) G \tl{G}, 
\end{align}
where the first two terms are the extra PQ breaking terms, and 
$c_g$ is the axion coupling to gluons which depends on 
the UV model. 
We have inserted a $\pi$ in the last term because of the sign convention of the first term.  With this notation, it is easy to see that the axion hilltop inflation is possible if  $\kappa$ and $\Theta$ are in the range of (\ref{cfei}).
{The price we have to pay is the tuning of the relative height and phase, i.e., $\kappa$ and $\Theta$, but such a tuning of the parameters is common to most small-field inflation models, and is not peculiar to the axion hilltop inflation.}

 In the previous example,  the first two cosine terms come from the small-size instantons of the extended QCD gauge groups. In this case, we have $c_g = 1 \pm n$, where we restrict $n > 1$. We note that $n$ must be an even integer (i.e., $c_g$ is an odd integer) 
 so that all the cosine terms have the minimum at $\phi/f_\phi \simeq \pi$, which is ensured by the
positive path-integral measure of QCD~\cite{Vafa:1984xg}.\footnote{W.Y. thanks Ryuichiro Kitano for discussing this point in a different project.}   On the other hand, it is possible that one of the first two terms (e.g. the second term) arises from another PQ breaking, which is not related to the extension of the QCD. In this case $c_g$ is not related to $n$, and $n$ can be an even or odd integer.
In order to set the strong CP phase $\bar{\theta}_{\rm QCD} = 0$,  $c_g$ must be an odd integer in this case, too.

As we have seen in the previous section, the axion hilltop inflation requires $\Theta \approx 0$. 
From \Eqs{feven}  or \eq{fodd}, the potential minimum is at $ \langle \phi \rangle /f_\f  \approx \pi$, and thus, $\bar{\theta}_{\rm QCD} \approx 0$ (mod $2\pi$).\footnote{On the other hand, if $c_g$ is an even integer, we have $\bar{\theta}_{\rm QCD}\approx \pi$, and the scenario may be excluded by the spectrum of hadron given the masses of the up and down quarks~\cite{Ubaldi:2008nf}}
See also the footnote \ref{ft2} for the case of the fractional $c_g$ and/or $n$.
Thus, the axion hilltop inflation is not merely feasible in the heavy QCD axion scenario.  In fact, requiring the inflation to occur in this minimal setup results in the strong CP phase that is very close to zero.
The vanishingly small strong CP phase is tightly related to the axion hilltop inflation, {through the fine-tuning of the parameters required for the slow-roll inflation.}
This is one of the novel features of this scenario.

In the remainder of this section, we discuss the phenomenological aspects of the heavy QCD axion inflation models for even and odd $n$ separately, based on the analysis in Refs.\,\cite{Daido:2017wwb,Daido:2017tbr,Takahashi:2019qmh}.
We show that the predicted relation between the axion mass and coupling to gluons will be covered by many future experiments 
for $m_\phi \lesssim {\cal O}(10)$\,GeV, and study the implications for nucleon EDM, and the constraints on the inflation scale from the QCD dynamics.

\subsection{QCD axion inflation with even $n$}
When $n$ is even,  we have \beq \bar{\theta}_{\rm QCD}\approx \Theta/2 =\O(f_\f^3/M_{\rm pl}^3)\laq{ThetaCP}\eeq from \Eq{feven}, which turns out to be too small to be observed. 
This is partly because both the inflation scale and the decay constant cannot be arbitrarily high.
The contribution of the small-size instantons is bounded above since the corresponding gauge coupling is related to the QCD one at the symmetry breaking scale. Thus, there is an upper bound on the inflation scale and the decay constant from a theoretical point of view.  
For instance, in the UV model given in Appendix \ref{chap:app1}, we obtain a low-energy effective theory with $\L \lesssim 10^{5}\GEV$ (or equivalently $f_\phi \lesssim 10^8 \GEV$), and
the axion contribution to the nucleon EDMs 
is much smaller than that from the CKM phase.  Also,  the inflationary Hubble parameter satisfies $H_{\rm inf}\ll \KEV$ (see the footnote \ref{ft1}).

In Fig.\ref{fig:2} we show the parameter region (red band) in the $m_\f-f_\f^{-1}$ plane predicted by the QCD axion inflation model with with $n=2$ and $c_g=1$~\cite{Takahashi:2019qmh}, together
with the expected reaches of various experiments such as CODEX, DUNE, FASER, LHC, MATHUSLA, and NA62\footnote{The projected reach of the NA62, which is from $K\to \pi +\phi$, 
may be too conservative, and it can be significantly improved~\cite{Gori:2020xvq, Bauer:2021wjo}.  We thank Kohsaku Tobioka for pointing this out to us. In addition, with $n> \O(10)$ the inflaton can be searched for from B meson decays at the Belle-II experiment~\cite{Bertholet:2021hjl}.} as well as the cosmological and astrophysical bounds, adopted from \cite{Kelly:2020dda}. See also Ref.~\cite{Agrawal:2021dbo} for certain theoretical uncertainties of the estimated experimental reaches.
We note that  the IR QCD instanton potential is not entirely negligible during inflation for $f_\f\lesssim 10^{6-7}\GEV$, but its effect can be absorbed by the redefinition of $\kappa$ and $\Lambda$, whose effect is negligible in this 
figure. Interestingly, the predicted relation will be covered by many experiments. Considering that the decay constant is bounded above as $f_\phi \lesssim 10^8 \GEV$ from a theoretical point of view,  we can see that most of the predicted parameter region can be probed. 
If we take $c_g>1$, the predicted region shown as the red band moves upward by a factor of $c_g.$ Thus our conclusion does not change for $c_g=\O(1)$. 
For $c_g \gg 1$, the parameter region can be tested by the heavy meson decay in the Belle II experiment~\cite{Chakraborty:2021wda}.

\begin{figure}[!t]
\begin{center}  
     \includegraphics
[width=145mm]{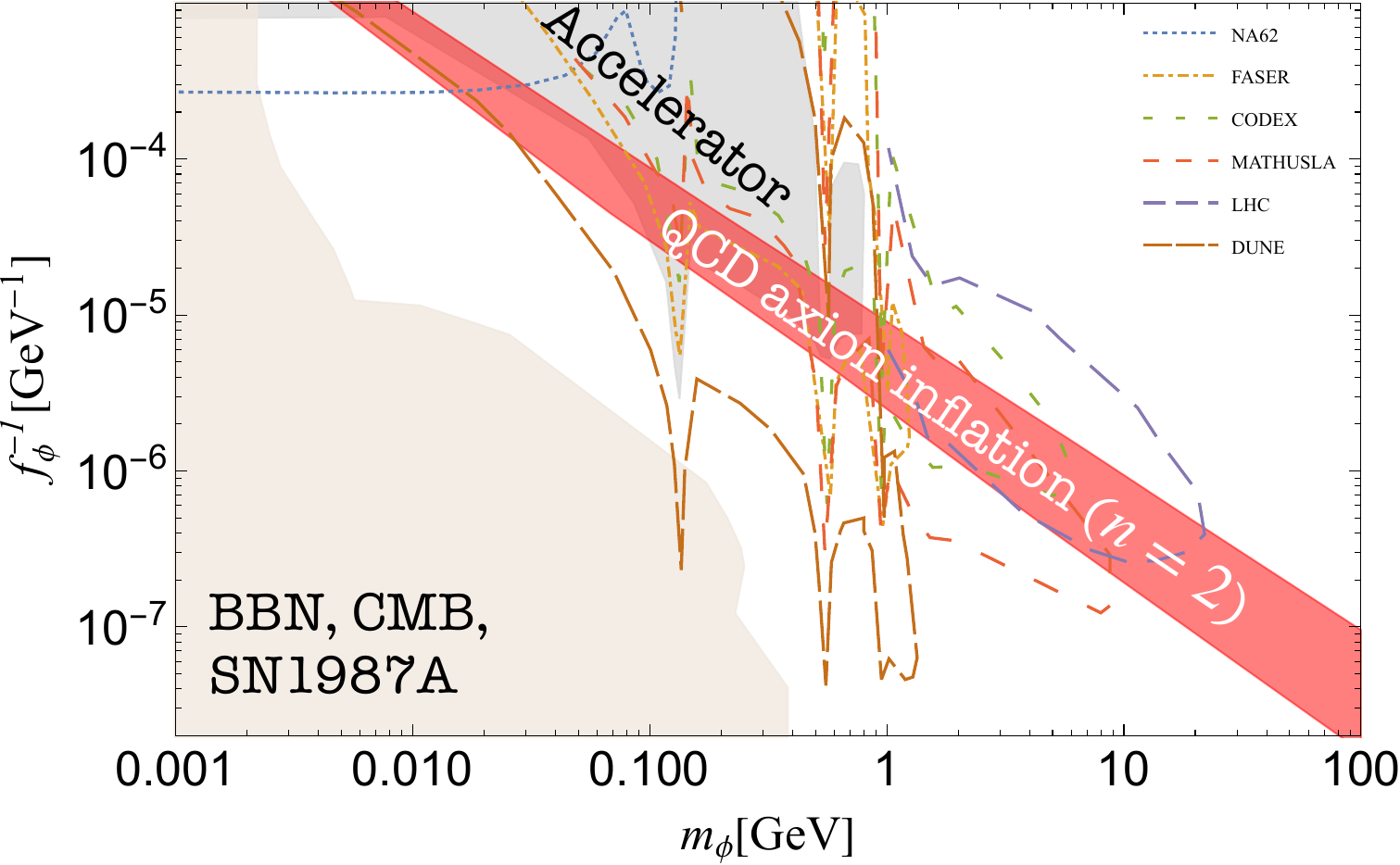}
      \end{center}
\caption{The predicted parameter region of the QCD axion inflation model with $n=2$ and $c_g=1$  shown as the red band. 
 Various bounds and reaches are adopted from Ref.\,\cite{Kelly:2020dda}. Note that $f_\phi$ is bounded above as $f_\phi \lesssim 10^8 \GEV$.
 }
\label{fig:2} 
\end{figure}

After inflation, the axion oscillates around the potential minimum, and forms a condensate. 
It decays via the same coupling to gluons,  scatters off with the decay products, and eventually quickly evaporates~\cite{Daido:2017wwb,Daido:2017tbr,Takahashi:2019qmh}. 
{The dominant process for reheating is the dissipation, whose rate is given as the larger of 
the QCD sphaleron contribution~\cite{McLerran:1990de, Moore:2010jd}:
}
\beq
\Gamma^{\rm sph}_{\rm dis} \sim  (3\a_s)^5\frac{T^3}{f_\phi^2},
\eeq
and the perturbative scattering~\cite{Yokoyama:2005dv,Anisimov:2008dz,Drewes:2010pf,Mukaida:2012qn,Drewes:2013iaa,Mukaida:2012bz, Moroi:2014mqa}:
\beq
\Gamma^{\rm pert}_{\rm dis} \sim  \frac{\a_s^2 T^3}{32\pi^2 f_\f^2}  \frac{m_\phi^2}{g_s^4 T^2}.
\eeq
{The axion condensate evaporates} and the 
reheating is successful if $H \lesssim \max {[\Gamma^{\rm sph}_{\rm dis}, \Gamma^{\rm pert}_{\rm dis}]}|_{T\sim \L}$.
For the parameters of our interest, the reheating is instantaneous and the reheating temperature is $T_R \sim \L.$ In particular, $T_R$ exceeds the weak scale $\sim 100 \GEV$ for  $m_\f \gtrsim 0.1\GEV$. In this case, the baryon asymmetry can be generated by the electroweak (EW) baryogenesis or resonant leptogenesis.

In this set-up the QCD axion is unstable and cannot be dark matter. One dark matter candidate is a light(er) axion. As mentioned in the footnote \ref{ft1}, the initial displacement of the light axions can be suppressed in the eternal inflation scenario. However, it is possible to excite those light axions by making use of the mixing with the inflaton~\cite{Daido:2017wwb,Takahashi:2019pqf,  Takahashi:2019qmh, Kobayashi:2019eyg, Nakagawa:2020eeg}. (See also Refs.~\cite{Co:2018mho,Huang:2020etx}.)

Another candidate is 
 a weakly coupled right-handed neutrino, which can be produced from the axion decay. 
In this case the right dark matter abundance can be explained with the dark matter mass satisfying~\cite{Moroi:2020has}
\beq
m_N\sim  0.4 \KEV \sqrt{C}c^{-1}_N
\frac{m_\f}{1\GEV},
\eeq
where $C$ and $c_N$ are the coefficient of the evaporation rate, and the PQ charge of the right-handed neutrino, respectively, and their natural values are of ${\cal O}(1)$.
Interestingly, the right-handed neutrino is cold
even though it is produced by the axion decay. 
This is because its initial momentum is of order the axion mass, which is much lower than the reheating temperature, i.e.  $\sim m_\f\ll T_R.$ 

We also note that the inflation scale must be very low since the small instanton contribution for the heavy QCD axion potential cannot be arbitrarily large.
The tensor to scalar ratio is predicted to be 
\beq
 r \simeq 2 \times 10^{-41} \left( \frac{{H_{\rm inf }}}{
     \hbox{1 keV}}\right)^2 ,
\eeq
which is far below the future sensitivity reach. 
Therefore, the future detection of the primordial gravitational wave from inflation can fully exclude this scenario. {This is also the case of odd $n$, which will be discussed soon.}

\subsection{QCD axion inflation with odd $n$}
When $n$ is odd,  the inflaton contribution to the strong CP phase can be sizable. From \Eqs{cmax} and \eq{fodd}, we have
\beq
\laq{oddphase}
\bar{\theta}_{\rm QCD}\simeq -\(\frac{6\x}{n^2-1}\)^{1/3} \frac{f_\f}{M_{\rm pl}}.
\eeq
The measurement of the neutron 
EDM~\cite{Baker:2006ts,Afach:2015sja} can be translated to the upper bound
$\ab{\bar{\theta}_{\rm
QCD}} \lesssim 2\times 10^{-10}$~\cite{Pospelov:2005pr, Dragos:2019oxn}. 
Thus, the axion decay constant is constrained to be
the following range,
\beq
10^8\GEV \lesssim f_\phi \lesssim 2.6\times 10^{9} \GEV \(\frac{10^{-3}}{\x /(n^2-1)}\)^{1/3},
\eeq
where  the (conservative) lower bound is from the duration of the neutrino burst from the SN1987A~\cite{Mayle:1987as,Raffelt:1987yt,Turner:1987by, Chang:2018rso,Carenza:2019pxu} and the cooling argument of neutron stars \cite{Hamaguchi:2018oqw,Leinson:2019cqv}.

We show in Fig.~\ref{fig:1} the predicted parameter region (blue band) for the case of $n=3$ 
in the $m_\f-f_\f^{-1}$ plane. Note that
the axion mass and decay constant are related to each other in this model, because the axion mass is comparable to the Hubble parameter during inflation.  The contours of $ |\bar{\theta}_{\rm QCD}| \times 10^{10}$
are also shown on the blue band.
Interestingly, the axion  mass is related to the EDM of nucleon, which is proportional to $|\bar{\theta}_{\rm QCD}|$.  The axion mass cannot exceed $\sim 3\KEV$ due to the too large contribution to the strong CP phase. 
The upper shaded region (purple) represents the SN1987A bound. The neutron-star cooling bounds~ \cite{Hamaguchi:2018oqw,Leinson:2019cqv} are denoted by the two horizontal dashed (green) lines

The lower bound of the decay constant restrict the $|\bar{\theta}_{\rm QCD}|\gtrsim 10^{-12}\text{-}10^{-11}.$
Future storage ring experiment is expected to improve the sensitivity on the proton EDM by three orders of magnitude~\cite{Anastassopoulos:2015ura,Omarov:2020kws}. Thus, it will fully cover the predicted parameter region,  if the proton EDM can be estimated precise enough from theoretical calculations.  
The neutron EDM bound also restricts the inflation scale from \eq{fodd} to be below $\KEV$, or equivalently,
 $m_\phi \lesssim 3\KEV$ (see the footnote \ref{ft1}).

\begin{figure}[!t]
\begin{center}  
     \includegraphics
[width=145mm]{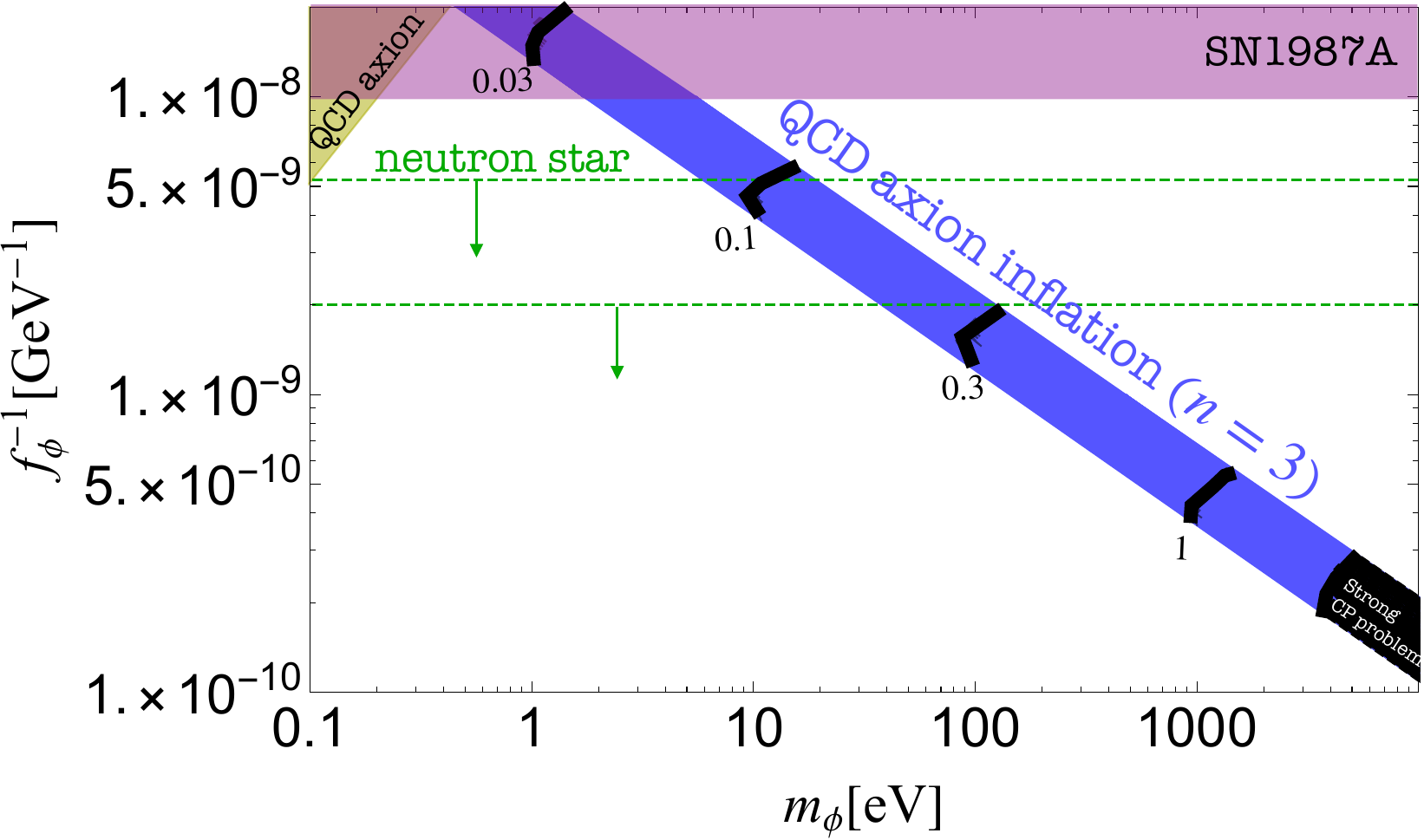}
      \end{center}
\caption{
The predicted parameter region of the QCD
axion inflation model with $n=3$ and $c_g=1$  shown as the blue band. 
The contours on the blue band denote the 
$|\bar{\theta}_{\rm QCD}| \times 10^{10}.$ The black shaded region is excluded due to too large neutron EDM. 
 The upper shaded (purple) region is disfavored by the duration of the neutrino burst from the SN1987A. The neutron star cooling bounds of  \cite{Leinson:2019cqv} and \cite{Hamaguchi:2018oqw} are shown by the top and bottom (green) dashed lines, respectively.  
For comparison the QCD axion prediction is shown in the shaded (yellow) region in the upper left.
}\label{fig:1} 
\end{figure}

In this model, however, 
 the decay constant is too large for the axion to reheat the Universe efficiently within  this effective theory.\footnote{If  the axion is coupled to other SM particles, the inflaton condensate can evaporate efficiently. However, thermally produced axions in the evaporation process would contribute too much to the hot or warm dark matter~\cite{Takahashi:2020bpq}. }
If we introduce another axion which has a mixing with the QCD axion, the end of inflation is like a hybrid inflation, and the subsequent reheating can be successful through the resonant conversion~\cite{Daido:2017wwb}.\footnote{
Hybrid inflation using axions was studied in e.g. Refs.~\cite{Peloso:2015dsa,Daido:2017wwb,Carta:2020oci}.
} 
Then, the QCD axion may remain as dark matter. 
However, if the axion-photon coupling is same as
the KSVZ axion~\cite{Kim:1979if,Shifman:1979if},
the axion dark matter in this mass range is either in tension with X-ray observation, extragalactic background light observation, or the reionization history (see e.g. Ref.~\cite{Arias:2012az}). 
 We may introduce additional couplings of the axion to $\U(1)_Y$ or $\SU(2)_L$ gauge fields to suppress the photon coupling by $\O(1\text{-}10)\%$ to relax the tension.

\section{Discussion}
\label{sec:4}

The question is how to get the parameter of $\k$ and $\Theta$ in the  range (\ref{cfei}) for the hilltop inflation to occur. 
{ 
In the heavy axion model discussed in the previous section, these conditions are nothing but fine-tuning, and this is the biggest theoretical difficulty with this heavy QCD axion inflation model. In terms of the degree of fine-tuning, since $\delta\kappa/\kappa
\sim \delta \alpha_i/\alpha_i$, the gauge coupling(s) needs to be fine-tuned to an accuracy of $(f_\phi/M_{\rm pl})^2$. Similarly, the relative phase of the corresponding potential terms needs to be adjusted to an accuracy of $(f_\phi/M_{\rm pl})^3$.
When compared to the degree of fine-tuning of the strong CP phase, if $f_\phi$ is smaller than $10^{15}$\,GeV, the degree of fine-tuning required for inflation is greater than the strong CP problem.
}

A possible explanation is the anthropic requirement for the slow-roll inflation.
{One reason why the strong CP problem is so serious is that, unlike the cosmological constant or the Higgs mass, the experimental upper bound on the strong CP phase is much smaller than the anthropic upper limit~\cite{Ubaldi:2008nf} (see also Refs.~\cite{Takahashi:2008pu,Kaloper:2017fsa,Dine:2018glh}). In our model,  the strong CP problem or the high quality problem of the PQ symemtry in the PQ mechanism is replaced with the fine-tuning of the parameters for the slow-roll inflation. As mentioned above, unless $f_\phi$ is as large as $10^{15}$\,GeV, the required fine-tuning becomes more severe, but it may
be interesting from the landscape perspective
that an anthropic explanation may work.
}

Another direction is to build a concrete UV set-up that realizes the conditions on the parameters, and in fact, such a  possibility was studied in certain UV models with extra-dimensions {in the context of multi-natural inflation}~\cite{Croon:2014dma,Higaki:2015kta, Higaki:2016ydn}.{It is worth studying the model building of the heavy QCD axion inflation along the same lines.}

One interesting possibility to reduce the amount of fine-tuning is to increase the inflation scale.\footnote{
We thank the anonymous referee for bringing this point to our attention.}
In the case of the small-size instanton considered in the previous section, there is an upper limit on the inflation scale, but if we consider another source of the PQ symmetry breaking, we can further increase the inflation scale. In this case, at least one potential term must be aligned with the potential from the IR QCD effects.
As an example, let us consider a mirror SM sector which is related to the SM  by $Z_2$ symmetry. We assume that the $Z_2$ symmetry is broken by a VEV of hidden Higgs, $v_H$, which is much larger than the EW scale.
The axion couplings to gluons and hidden gluons (shown with primes) are
\beq
{\cal L} = \frac{\a_s}{8\pi}( \frac{\f }{f_\f} -\theta_{CP}) G\tl G +\frac{\a'_s}{8\pi}( \frac{\f }{f_\f} -\theta_{CP}) G'\tl G'. 
\eeq 
at a renormalization scale, $\m_{\rm RG} (<v_H)$,
where both sectors are still perturbative.
In this setup, while  $\a_s' \sim \a_s$ at $\mu_{\rm RG}\sim v_H$, 
 we have $\a_s'> \a_s$ at $\mu_{\rm RG} \ll v_H$
because of decoupling of the heavy colored particles in the hidden sector. The hierarchy between  $\a_s'$ and $\a_s$ can be even larger if there are extra
colored particles in both sectors, and they get masses $v_H$ in the hidden sector while their $Z_2$ partners
in the SM sector remain relatively light.
Then, the axion receives a large potential term from non-perturbative effects of the hidden QCD, \cite{Fukuda:2015ana}
\beq
V\sim 
\Lambda'^4
\cos {(\f/f_\f-\theta_{CP})},
\eeq
whose minimum aligns with the potential from the IR QCD effect. 
If we require $f_\phi \gtrsim 10^{15} \GEV$ to relax the fine-tuning for successful inflation, we need to have $\Lambda' \gtrsim 10^{13}\,\GEV$
to satisfy the CMB normalization condition. However we also notice that, once this inequality is satisfied, the successful inflation does not explain the smallness of the strong CP phase, and we need an extra tuning to make $\bar \theta_{\rm QCD}$ smaller than $10^{-10}$ (see \Eqs{ThetaCP} and \eq{oddphase}). Thus, the fine-tuning of the parameters as a whole cannot be relaxed more than this. In the marginal case of $f_\phi \sim 10^{15}\,\GEV, \Lambda'\sim 10^{13}\,\GEV$ with even $n$, the nucleon EDM may be tested in the future.
By introducing another cosine potential for $\f$,  we obtain the same potential of \Eq{DIV}. If $\L'$ is large enough, the fine-tuning of the parameters can be ameliorated. 
However, we also expect other operators which breaks the alignment of the minima. For instance, we may have
$(|H_{\rm hidden}|^2-|H_{\rm SM}|^2)(G'\tl G'-G\tl G )/M^2_{\rm pl} \supset \frac{v_H^2}{M^2_{\rm pl}} G'\tl G'$,
where $H_{\rm hidden}\AND H_{\rm SM}$ are the hidden and SM Higgs fields.
Then, the the inflation model can solve the strong CP problem if $v_H \lesssim 10^{13}\GEV$.\footnote{Note that, in the presence of a gauge singlet scalar whose VEV breaks the $Z_2$ symmetry, the bound
can be much tighter. If we have a term like
 $\frac{v_H}{M_{\rm pl}} (G'\tl G'- G \tl G)$,
where the coefficient $v_H$ arises from a VEV of another $Z_2$ odd real singlet scalar~\cite{Fukuda:2015ana}, we need to have $v_H/M_{\rm pl} \ll 10^{-10}$.}

Lastly let us discuss to what extent the heavy QCD axion inflation works in the presence of other PQ symmetry breaking terms.
So far we have assumed that the QCD axion potential receives two instanton contributions which  increase the axion mass compared to the usual QCD axion mass.
Given that the natural inflation with a single cosine term has been already excluded by observation, 
the axion potential considered in this letter is its simplest and the minimal extension.
However, there may be other PQ symmetry breaking terms which change the inflaton dynamics and/or contribute to the strong CP phase. It is possible to realize successful slow-roll inflation in the presence of more than two instanton contributions, but the relation between the mass and decay constant can be modified especially in the odd $n$ case. This is because the light axion mass at the minimum is due to the upside-down symmetry of the potential, which is generically lost in the presence of other (sizable) PQ breaking terms. Let us estimate on the conservative condition on such an extra PQ breaking term.  
Suppose that the dangerous PQ breaking terms have the form of $ \frac{f_\f^{4+r}}{M_{\rm pl}^r} \cos[\frac{c_r \f}{f_\f}+ \theta_r]\sim  c_r \theta_r  \frac{f_\f^{4+r}}{M_{\rm pl}^r}\frac{ \f}{f_\f},$ 
where we have expanded the cosine-term around $\f\sim 0$ by assuming $|\theta_r| \lesssim 1$ in the last approximation. 
Let us require that the axion potential during inflation is not disturbed by this extra term.
The most sensitive part in  \Eq{app} is the linear term proportional to $\Theta \sim f_\f^3/M_{\rm pl}^3.$ By requiring the PQ breaking term to be smaller than the linear term in \eq{app}, the inflation dynamics remains unchanged if 
$f_\f\lesssim 10^{-\frac{12}{r-3}} \times M_{\rm pl} $
for $r>3.$ 
If this is satisfied, the inflation dynamics is not modified, and the contribution to the strong CP phase is negligible. 
For the parameter region of interest, the term with $r\geq 5$ can be safely neglected.
The situation is somewhat analogous to the PQ quality problem; for the PQ mechanism to solve the strong CP problem, the PQ symmetry must be predominantly broken by the IR non-perturbative QCD effect, and other possible breaking terms must be extremely suppressed unless their minima are aligned with $\bar{\theta}_{\rm QCD} =0$. 
{In ordinary heavy QCD axion scenarios, such quality problem can be relaxed if the decay constant is small enough. In our scenario, on the other hand, 
the quality problem is closely related with the slow-roll conditions for successful inflation.  }

\section{Conclusions}

{In this paper, we have studied whether the heavy QCD axion can drive inflaton while solving  the strong CP problem.  We have shown that the axion hilltop inflation can be implemented in the heavy QCD axion scenario at the expense of fine-tuning parameters involved. To this end, the relative height and phase of the two instanton contributions must be aligned such that the QCD axion potential is very flat around the maximum, which enables the slow-roll inflation. Interestingly, requiring the axion hilltop inflation results in the strong CP phase that is close to zero. Thus, the solution to the strong CP problem is closely related to the axion hilltop inflation in the minimal set-up with the two extra instanton contributions.  In our scenario, the degree of fine tuning the parameters required for slow-roll inflation is greater than the strong CP problem, and thus, it is fair to say that the fine-tuning gets worse by implementing the inflation in the heavy QCD axion sector. On the other hand, we may now be able to explain the small strong CP phase by an anthropic argument 
for inflation. If eternal inflation by the heavy QCD axion occurs in the landscape, a universe with a small strong CP phase may be realized as a consequence. Also, in this case, there is a relationship between the mass of the axion and the decay constant, and some of the parameter regions can be probed by future experiments.
}

\section*{Acknowledgments}
We thank Kohsaku Tobioka for useful comments.  
This work is supported by JSPS KAKENHI Grant Numbers 17H02878 (F.T.), 20H01894 (F.T.), and 20H05851 (F.T. and W.Y.).

\appendix 
\section{Small-size instantons of the extended QCD gauge groups}
\lac{app1}
In the main text we study the axion hilltop inflation with two instanton contributions. Here we discuss the case in which both arise from small-size instantons of the extended gauge groups of QCD. 

Let us study a concrete UV model with $\SU(3)^N\to \SU(3)_c$, similar to that proposed in Ref.\,\cite{Agrawal:2017ksf} for the heavy QCD axion. See also \cite{Csaki:2019vte} for the full one-instanton calculation. 
The difference from the reference's model is that we reduce the number of axions. 

Let us begin with the case of   $N=2$, i.e. $\SU(3)_1 \times \SU(3)_2  \to \SU(3)_c$. 
The axion couplings in the UV are given by
\beq
{\cal L} \supset \frac{\a_1}{8\pi}\( c_1  \frac{\f}{f_\f}-\bar \theta_1\) G_1 \tl{G_1}+\frac{\a_2}{8\pi}\( c_2  \frac{\f}{f_\f}-\bar \theta_2\) G_2 \tl{G_2}.
\eeq
After the Higgsing by a bi-fundamental Higgs field and integrating out the heavy degrees of freedom, 
we obtain the low-energy effective Lagrangian (c.f. \cite{Agrawal:2017ksf})
\beq
\laq{Lageff}
{\cal L}_{\rm eff}=  \L_1^4  \cos\(c_1\frac{\f}{f_\f} -\bar \theta_1\)+\L_2^4 \cos\(c_2\frac{\f}{f_\f} -\bar \theta_2\)+\frac{\a_s}{8\pi}\(( c_1 +c_2 ) \frac{\f}{f_\f}-\bar \theta_1-\bar \theta_2\) G \tl{G},
\eeq
Here the first two terms are from  the small-size instantons.
If we do not consider the axion hilltop inflation, the model cannot solve the strong CP problem unless we tune $\bar \h_1/c_1\approx \bar \h_2/c_2 \mod  \pi $. 

\subsection{Even $n$ case}
The first two terms can induce the axion inflation if the potential has a flat plateau.  The inflaton potential \Eq{Lag} is reproduced by rewriting $\L_1=\L, \L_2^4=\k \L^4/n^2$, $c_1 \f/f_\f -\bar{\theta}_1= \f/f_\f -\Theta-\pi$ and $c_2 \f/f_\f-\bar{\theta}_2= \pm n \f/f_\f.$ 
In this case the axion coupling to gluons is given by
 \beq 
 \label{cg1n}
 c_g=1\pm n.
 \eeq
 So, when $n$ is even, $c_g$ is odd. In this case, requiring the axion hilltop inflation results in $\theta_{\rm QCD} \approx 0$, solving the strong CP problem.
 
As noted in \cite{Agrawal:2017ksf,Csaki:2019vte},
the small instanton contributions to the QCD axion mass are not significant in the case of  $N=2$, but they can be much more enhanced in the case of 
$N\geq 3$.
For $N=3$, we have
\beq
\laq{UV1}
{\cal L} \supset \frac{\a_1}{8\pi}\( c_1  \frac{\f}{f_\f}-\bar \theta_1\) G_1 \tl{G_1}+\frac{\a_2}{8\pi}\( c_2  \frac{\f}{f_\f}-\bar \theta_2\) G_2 \tl{G_2}+\frac{\a_3}{8\pi}\( c_3  \frac{\f}{f_\f}-\bar \theta_3+ c_H \frac{\f_H}{f_H}\) G_3 \tl{G_3}
\eeq
where we put an additional axion, $\f_H$ with the decay constant $f_H$.
After the Higgsing we obtain the following additional terms 
\beq \d {\cal L}_{\rm eff}= -{\L}_3^4 \cos{\(c_3 \frac{\f-\f_{\rm min}}{f_\f} +c_H\frac{\f_H}{f_H}\)} - \frac{\a_s}{8\pi}\(c_3 \frac{\f-\f_{\rm min}}{f_\f} +c_H\frac{\f_H}{f_H}\) G \tl{G}
\eeq
where we have absorbed  $\bar{\theta}_3$ by a field redefinition of $\f_H$  and   $\L_3$ is the the resulting potential scale from small instanton. 
If $c_3$ is nonzero, the first term induces a mixing between the axions, which can modify the end of the inflation like hybrid inflation~\cite{Daido:2017wwb}.
We can integrate out the heavy axion during inflation
if \beq \laq{condH} H_{\rm inf} \ll \frac{\L_3^2}{f_H}.\eeq
Then we obtain the same effective Lagrangian \eq{Lageff},
but the inflation scale can be much larger than the case of $N=2$. 

After inflation, the reheating is instantaneous
for the parameters of our interest.
At the end of the reheating, the typical energy of the plasma scattering is $T_R \sim  \L \sim \L_1 \sim \L_2.$ 
For the effective theory for a single axion to be valid not only during inflation but also during the reheating process, therefore, we also need 
\beq 
T_R \ll \frac{\L_3^2}{f_H} \lesssim \Lambda_3,
\eeq
where the second inequality is due to the perturbative bound. 
According to Refs.\,\cite{Agrawal:2017ksf, Csaki:2019vte}, a larger $\L \sim \L_1 \sim \L_2$ is obtained with a smaller $\L_3.$ 
Therefore, by taking $\L_3 \sim \L_1 \sim \L_2$ and $f_H\sim \L_3$, 
we get the largest $\L$ for the effective theory,
and it is given by \cite{Agrawal:2017ksf, Csaki:2019vte} 
 \beq \laq{Lmax}\L_{\rm max}\sim 10^{4-5}\GEV.\eeq 
With $\L \lesssim \L_{\rm max}$ and by taking certain $f_H$ and $ \L_3$, we can obtain the model in the main text without introducing any other fields in the low energy. If, on the other hand, $\L_3^2/f_H$ is smaller than $T_R$, $\f_H$ may be produced due to the thermal scattering via the gluon coupling. Also it may be produced via the mixing with $\f$ if $H_{\rm inf}\lesssim \L_3^2/f_H \lesssim \L^2/f_\f.$  If $\L_3^2/f_H\lesssim H_{\rm inf}$ it may not be set at the potential minimum during and soon after the inflation. 
In either case, our discussion in the main part should be modified.

\subsection{Odd $n$ case}
When $n$ is odd, $c_g$ is even according to Eq.~(\ref{cg1n}).
This results in $\bar{\theta}_{\rm QCD}\approx \pi \mod 2\pi$ at the potential minimum.
This can be consistent with the neutron EDM bound since the vacuum at ${\theta}_{\rm QCD} = \pi$ is CP-conserving (by assuming no spontaneous CP breaking). 
However, it may be in tension with the hadron spectrum~\cite{Ubaldi:2008nf}. Whether we can get a consistent hadronic picture by changing the QCD parameters such as light quark  masses are beyond 
the scope of this letter.

We may introduce an additional heavy axion  $\f_{H_2}$ with the Lagrangian
\beq
{\cal L} \supset \frac{\a_1}{8\pi}\( c_1  \frac{\f}{f_\f}-\bar \theta_1\) G_1 \tl{G_1}+\frac{\a_2}{8\pi}\( c_2  \frac{\f}{f_\f}-\bar \theta_2+ c_{H_2} \frac{\f_{H_2}}{f_{H_2}}\) G_2 \tl{G_2}+\frac{\a_3}{8\pi}\( c_3  \frac{\f}{f_\f}-\bar \theta_3+ c_H \frac{\f_H}{f_H}\) G_3 \tl{G_3},
\eeq
and an additional cosine potential (with odd $n$) for $\phi$ to have a sufficient flat plateau as discussed in the main part. 
During inflation we can integrate out $\f_H$ and $\f_{H_2}$ if they are heavier than $H_{\rm inf}$. 

However, it is difficult to integrate out $\f_{H_2}$ and $\f_H$ , then  $\f_{H_2}$ and $\f_{H}$ may be produced after inflation, and their decay may reheat the Universe.

\bibliography{reference}
\end{document}